\documentclass[reprint,showpacs,amsmath,amssymb,aps,prb]{revtex4-1}
\usepackage{graphicx}
\usepackage{subfigure}
\usepackage{dcolumn}
\usepackage{bm}
\usepackage{verbatim} 

\usepackage{amssymb}

\begin{document}

\preprint{APS/123-QED}

\title{Waiting Time Distribution of Quantum Electronic Transport in Transient Regime}

\author{Gao-Min Tang}
\author{Fuming Xu}
\author{Jian Wang}
\email{jianwang@hku.hk}
\affiliation{Department of Physics and the Center of Theoretical and Computational Physics, The University of Hong Kong, Pokfulam Road, Hong Kong, China}

\date{\today}

\begin{abstract}
Waiting time is an important transport 
quantity that is complementary to average current and its fluctuation. So far all the studies of waiting time distribution (WTD) are limited to steady state transport (either dc or ac). The existing theory can not deal with WTD in the transient regime. In this regard, we develop a theoretical formalism based on Keldysh non-equilibrium Green's functions formalism to study WTD. This theory is suitable for dc, ac, and transient transport and can be used for first principles calculation on realistic systems. We apply this theory to a quantum dot system with a upward bias pulse and calculate cumulants of transferred charge as well as WTD in the transient regime. The oscillatory behavior of WTD is found in the transient regime. We give a general relation between WTD and experimental measured quantity and demonstrate its feasibility for a quantum dot system in the transient regime.
\end{abstract}

\maketitle

\section{Introduction}

Transport processes in mesoscopic systems are dominated by quantum effect and are stochastic in nature.\cite{kampen} Therefore in addition to average current, full probability distribution of charge transport called full counting statistics (FCS) is needed to fully characterize the quantum transport.\cite{lev1,lev2,lev3} Indeed, noise spectrum (the second cumulant of current operator) and higher order fluctuations can provide additional information about quantum effect and nature of the interaction in electronic systems.\cite{blanter,sam,klich,lesovik} Experimentally, high order transient cumulants of charge passing through a quantum point contact have been measured up to 15th cumulant and universal oscillations were found in counting statistics.\cite{flindt} Theoretically, finite frequency FCS has been studied using quantum master equation and scattering matrix approach.\cite{fazio,brandes1,flindt2,buttiker} Less attention has been paid on FCS of transient transferred charge.\cite{foot2,25}

Besides FCS, another complementary quantity to characterize the stochastic processes is the waiting time distribution (WTD) which is the distribution of time interval between two successive events.\cite{albert1} This quantity has been extensively studied in quantum optics\cite{11,14} long time ago which can provide us new insight of quantum correlation on the short time scale. Recently a scattering quantum theory for WTD was formulated by Albert et al that can be used to study WTD of dc electronic quantum transport\cite{albert2,rajabi} and has been extended to steady state ac regime.\cite{flindt3} Despite of the success of this quantum theory, there are many open questions remained to be answered. For instance, since this quantum theory is applicable only at zero temperature and can not be used for transient dynamics, it is clearly desirable to develop a new theory at finite temperatures and in the transient regime so that the fluctuation theory\cite{forster} can be discussed and switching dynamics can be studied. We notice that charge distribution function $P(n,t)$ has been measured experimentally for a quantum dot system in the Coulomb blockade regime\cite{flindt} from which WTD can be deduced. Beyond the Coulomb blockade regime, it is very difficult to measure the distribution function and hence the WTD. The interesting questions are how to calculate high order cumulants of transferred charge in transient regime in order to compare with experimental results? How to relate WTD to experimental measured quantity if WTD can not be measured directly? It is the purpose of this paper to address these questions.

The main issue of FCS is how to calculate the generating function (GF) or cumulant generating function (CGF) from which we can calculate higher cumulants, the probability distribution $P(n,t)$ and the WTD. Levitov and Lesovik have presented an analytical expression for the GF in the long-time limit using a gedanken experiment scheme of a "charge counter" in the form of spin precession.\cite{lev1,lev2,lev3} The theory of GF for current was generalized to a general quantum mechanical variable by Nazarov and Kindermann,\cite{6} and was extended to study short time behavior of dc and ac current using the wave-packet formalism.\cite{7,albert2,flindt3} In this paper, we develop a theoretical formalism for WTD for coherent conductors. Using the non-equilibrium Green's function\cite{28,38} and path integral method in the two-time quantum measurement scheme\cite{21}, we obtain GF for the electron transport system which allows us to study FCS and WTD in dc, ac, and transient regimes. As an application of this theory, we calculate the cumulants of transferred charge from the GF and WTD of a quantum dot coupled by two leads in the transient regime and examine their temperature dependent behaviors.\cite{foot1} Analytic results of very short and long time behaviors of WTD are obtained. In addition, we discuss how to obtain WTD from cumulants of transferred charge which have been measured experimentally. Finally we note that this general framework of NEGF can be combined with the density functional theory to study FCS and WTD from first principles.\cite{hong}

\section{Theoretical formalism}

The central idea of FCS is to derive the probability distribution $P(\Delta n,t_0,t)$ of the number of the transferred electrons $\Delta n=n_t-n_0$ between an initial time $t_0$ and a later time $t$ which can be done using two-time quantum measurement.\cite{21} Defining the Fourier transform of the probability distribution as the generating function (GF) or characteristic function, we have
\begin{equation} \label{eq1}
Z(\lambda,t_0,t)\equiv\left< e^{i\lambda\Delta n}\right>=\sum_{\Delta n}P(\Delta n,t_0,t)e^{i\lambda \Delta n} ,
\end{equation}
where $\lambda$ is the counting field and $\Delta n$ can be either positive or negative. The $j$th moment of transferred charge $\left<(\Delta n)^j\right>$ and the $j$th cumulant $\langle\langle(\Delta n)^j\rangle\rangle$ are given by:
\begin{equation} \label{eq3}
\left<(\Delta n)^j\right>=\frac{\partial^j Z(\lambda)}{\partial(i\lambda)^j} \bigg|_{\lambda=0} ,~~~~
\langle\langle(\Delta n)^j\rangle\rangle=\frac{\partial^j \ln Z(\lambda)}{\partial(i\lambda)^j} \bigg|_{\lambda=0}
\end{equation}
From the GF, the distribution function for the number of the electrons can be found
\begin{equation}     \label{eq4}
P(\Delta n,t_0,t)=\int_0^{2\pi}\frac{d\lambda}{2\pi}Z(\lambda,t_0,t)e^{-i\lambda \Delta n}
\end{equation}
In particular, the probability of no electrons detected during time t denoted as $P(0,t_0,t)$ (also called idle time probability) is found to be\cite{kampen,albert2}
\begin{equation}  \label{eq5}
\Pi(t_0,t)=P(0,t_0,t)=\int_0^{2\pi}\frac{d\lambda}{2\pi} Z(\lambda,t_0,t).
\end{equation}
Now we consider the WTD. In the steady state, if we detect an electron at $t=t_0$, the probability of finding the next electron at time $t_0+t$ is the waiting time which is related to the idle time probability. In this regime, WTD was found to be\cite{albert2}
$W(t) =\langle t \rangle \frac{d^2 \Pi(t)}{dt^2} $
where $\langle t \rangle$ is the average waiting time. We note that WTD depends only on t due to the translational symmetry on time in dc case. In the case of ac bias, averaging over a period was carried out so that WTD depends only on t again.\cite{flindt3} For the transient process, time translational symmetry is broken and there is no time periodicity either in the transport. Hence WTD depends on two time indices. WTD can be defined in two ways depending on the number of measurements performed.\cite{kampen} One can define it similar to Ref.\onlinecite{albert2} where two measurements were done at $t_0$ and $t_0+t$, respectively. We then have $W_2(t_0,t) = -\partial_{t_0} \partial_t \Pi(t_0,t)/f_1(t_0)$ where the subscript 2 denotes the number of measurement and $f_1(t_0)$ is a normalization factor. For transient dynamics the bias is turned on at $t=0$, naturally we set $t_0=0$ in calculating $W_2$. Since the probability of finding electron is zero at $t=0$, it is not necessary to perform the first measurement at $t=0$. Instead, we ask if we start observation at $t=0$ how long we have to wait for the detection of an electron. This is the second definition of WTD $W_1$ with only one measurement. Obviously we have $W_1(t)=0$ at $t=0^+$ and $t=+\infty$. Since $\int_0^{t}W_1(\tau)d\tau$ is the probability of finding electrons in time interval $t$, the idle time probability $\Pi(t)$ satisfies $1-\int_0^{t}W_1(\tau)d\tau=\Pi (t)$. Hence the WTD for transient processes can be expressed by the idle time probability as
\begin{equation} \label{eq6}
W_1(t)=-\frac{d}{dt}\Pi(t) .
\end{equation}

\subsection{Generating Function}
To calculate the GF, we consider an isolated quantum dot and with two semi-infinite leads. The couplings between the two leads and the quantum dot are switched on at $t=0$ so that $\rho(0^-)=\rho_L\otimes\rho_D\otimes\rho_R$, where we have used L, R and D to denote the left and right lead as well as the quantum dot, respectively. In addition, a step-like pulse is also applied to the left lead at $t=0$. Note that the coupling between leads and the quantum dot and the external bias are turned on at the same time $t=0$. This transient problem that is slightly different from the Cini's approach (partition free approach) where the coupling between leads and scattering region are turned on in the remote past while the bias is turned on at $t=0$.\cite{cini}

Using the path integral formalism\cite{38}, the GF based on two-time quantum measurement approach can be expressed in terms of Keldysh nonequilibrium Green's function which is given by\cite{foot4}
\begin{equation} \label{eq8}
Z(\lambda,t_1,t)=\det (G \widetilde{G}^{-1})
\end{equation}
where,
\begin{equation}  \label{eq9}
G^{-1}=g^{-1}-\Sigma_L-\Sigma_R, ~~~ \widetilde{G}^{-1}=g^{-1}-\widetilde{\Sigma}_L-\Sigma_R .
\end{equation}
Here $\widetilde{\Sigma}_L$ denotes the self-energy containing the counting field and $g$ is the Green's function of the isolated quantum dot. Note that the counting field is between $t_1$ and t while the bias is turned on at time $t=0$. The Green's functions $G$ and $g$ as well as the self-energies $\Sigma_L$ and $\Sigma_R$ are all defined in the Keldysh space with the complex time contour being defined from time $t=0$ to time $t$ and then back to $t=0$. Hence the determinant has to be evaluated in Keldysh space whose dimension is $t$. In the Keldysh space the Green's function and self-energy have the following form:
\begin{equation} \label{eq10}
A(\tau,\tau^\prime)
= \left(
  \begin{array}{cc}
    A^r(\tau,\tau^\prime)   &   A^k(\tau,\tau^\prime)  \\
    		0   			 &   A^a(\tau,\tau^\prime)  \\
  \end{array}
\right)
\end{equation}
where $A^k=2A^<+A^r-A^a$. Finally the self-energy $\widetilde{\Sigma}_L$ in Eq.~(\ref{eq9}) is defined as
\begin{equation}  \label{eq11}
\widetilde{\Sigma}_L(\tau,\tau')= \Lambda^*(\tau) \Sigma_L(\tau,\tau') \Lambda(\tau')
\end{equation}
where
$\Lambda(\tau)=\exp[-i(\sigma_x +I)\lambda/2]\theta(\tau-t_1)$. We see that in the limit $t_1 \rightarrow \infty$ while keeping ${\bar t}=t-t_1$ finite, this formalism recovers the generating function $Z(\lambda, {\bar t})$ of dc transport.\cite{21}

Using Eq.(\ref{eq8}), both $W_2(t_1,t)$ and $W_1(t)$ can be investigated. Since $W_2(t_1,t)$ is much more complicated and computational more demanding numerically, we will focus in this paper on investigating $W_1$ and related quantities in detail. In this case, $t_1=0$ and we will drop the first time index $t_1$ from now on. Using Eq.(\ref{eq11}) the GF in (\ref{eq8}) can be written in the following form:
\begin{equation}  \label{eq12}
Z(\lambda ,t)=\det[I-G(\widetilde{\Sigma}_L -\Sigma_L)]
=\det [I-G M (e^{-i\sigma_x \lambda}-I)]
\end{equation}
where I is identity matrix and M is given by
\begin{equation} \label{eq13}
M(\tau,\tau^\prime)
= \frac{1}{2}\left(
  \begin{array}{cc}
    -\Sigma^a_L +\Sigma^r_L    &   	\Sigma_L^k  \\
    		-\Sigma_L^k   & 	 \Sigma^a_L -\Sigma^r_L   \\
  \end{array}
\right)_{(\tau,\tau^\prime)}
\end{equation}

In order to get various cumulants from Eq.~(\ref{eq3}), we take the derivative of the CGF which is $\ln Z(\lambda,t)$ with respect to $\lambda$
The transferred charge during time $t$ is:
\begin{equation} \label{eq14}
\langle \Delta n \rangle
= {\rm Tr}[(G^r-G^a) \Sigma^<_L + G^< (\Sigma^a_L-\Sigma^r_L)]
\end{equation}
where the trace is over both time space and real space. The current is obtained by taking time derivative of transferred charge,
\begin{align} \label{eq15}
I(t) = \int_{0}^{t}d\tau {\rm Tr}[G^r(t,\tau)\Sigma_L^<(\tau,t)+G^<(t,\tau)\Sigma_L^a(\tau,t)]+h.c..
\end{align}
Higher order cumulant of charge transfer can be calculated using Eq.~(\ref{eq3}). For instance, we find the charge-charge correlation to be
\begin{equation} \label{eq16}
\langle\langle (\Delta n)^2 \rangle \rangle= -{\rm Tr}[(G M \sigma_x)^2+GM].
\end{equation}

\bigskip

\subsection{Short and long times behaviors}
The WTD can be calculated using Eqs.(\ref{eq5}) and (\ref{eq6}). Now we examine its very short and very long time behaviors. Since ${\rm Tr}[GM]$ is proportional to $t^2$ as $t$ goes to zero, we find from Eq.(\ref{eq12}) $Z = 1- {\rm Tr} [GM (e^{-i\sigma_x \lambda} -1)]$ where we have used the relation ${\rm Det}(B) = \exp[{\rm Tr}\ln B]$. This in turn gives $P(n,t)=\delta_{n,0}-(1/2){\rm Tr}(GM) (\delta_{n,1}+\delta_{n,-1}-2\delta_{n,0})+(1/2)\langle \Delta n \rangle (\delta_{n,1}-\delta_{n,-1})$ by averaging over the counting field, where $\langle \Delta n \rangle$ is given by Eq.(\ref{eq14}). We find the idle time probability $\Pi(t) = 1+{\rm Tr} (GM)$ and distribution function $P(\pm 1,t)=(1/2)[-{\rm Tr} (GM)\pm \langle \Delta n \rangle]$. In general we have $P(n,t) \sim t^{2|n|}$ at short times. Hence the probability of finding two or more electrons is zero up to $t^2$. Finally, we arrive at the short time behavior of WTD
\begin{equation}
W_1(t)=2\int^t_0 d\tau {\rm Tr}[2G^k(t,\tau)\Sigma^<(\tau,t)+ G^<(t,\tau)(\Sigma_L^r(\tau,t)-\Sigma_L^a(\tau,t))]. \nonumber
\end{equation}
Obviously $W_1(t)$ is linear in $t$ for very small $t$. Our numerical result confirms this behavior. It is easy to show that the next order contribution to WTD is of the third order in $t$.

At very long time and zero temperature, we have
\begin{equation} \label{eq18}
\ln Z = t\int_0^{\Delta_L} \frac{dE}{2\pi}{\rm Tr} \ln[1+T(E) (e^{i\lambda}-1)]
\end{equation}
where $T(E)=\Gamma_L G^r \Gamma_R G^a$ is the transmission matrix and $\Delta_L$ is the Fermi level of the left lead. Obviously, Eq.(\ref{eq18}) gives $P(-n,t)=0$ for $n>0$ which is expected since at long times there is no electron going to the left. Taking selective discrete time $t$ as an integer $t_m$ and expanding GF in powers of $e^{i\lambda}$, we have
\begin{eqnarray} \label{eq19}
Z(\lambda)& =&\Pi_E ([1+ T (e^{i\lambda}-1)]^{t_m}) \nonumber \\
&\approx& e^{-\kappa t_m}[1+ a t_m e^{i\lambda}]
\end{eqnarray}
where $\Pi_E$ stands for multiplication over energy, $\kappa = -\int_0^{\Delta_L} (dE/2\pi) {\rm Tr}[\ln (1- T(E))]$ and $a=\int_0^{\Delta_L}( dE/2\pi) {\rm Tr}[T/(1-T)]$. After integrating $\lambda$ from 0 to $2\pi$, we find
\begin{equation} \label{eq20}
P(1,t)=a t e^{-\kappa t},~~~~ \Pi(t)=e^{-\kappa t},~~~~ W_1(t)=\kappa e^{-\kappa t}
\end{equation}
So the long time behavior of WTD is Poissonian as expected.

\bigskip

\section{Numerical Results}
We now apply our theory to a simple quantum dot connected by two leads. since the electronic structure of the leads can be important, we abandon the wideband limit and consider lead with finite band width\cite{29} ${\bf \Gamma}_{\alpha}(\epsilon)=\frac{\Gamma_{\alpha} W_0^2}{\epsilon^2+W_0^2} $
where $\alpha$ stands for the left or right lead, $\Gamma_{\alpha}$ is the linewidth amplitude and $W_0$ is the bandwidth, and assume that $\Gamma_L=\Gamma_R=\Gamma/2$. In the calculation we take $\Gamma$ as the energy unit and hence the time and current are measured $1/\Gamma$ and $e\Gamma$, respectively. In this paper, we choose bandwidth to be $W_0=10\Gamma$, the energy level of the quantum dot as $\epsilon_0=5\Gamma$, the Fermi levels of the left and right leads to be zero initially at $t=0^-$. We change the Fermi level of the left lead at $t=0$ to be $\Delta_L=10\Gamma$. Since the determinant of Eq.~(\ref{eq8}) is in the time domain, we have to calculate all the Green's functions and self-energies in the time domain which is given in Supplemental material.\cite{supp} In the following we present results of the idle time probability $P(0,t)$ and the probability for detecting one electron either from the left $P(1,t)$ or from the right $P(-1,t)$ during the time interval $t$ by integrating Eq.~(\ref{eq4}) numerically.

\begin{figure}
  \includegraphics[width=3.4in]{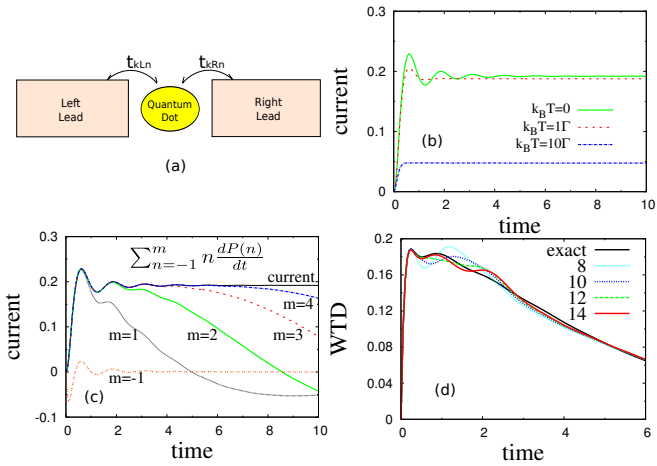}\\
  \caption{Transient current and WTD. (a). A schematic plot of the setup. (b). transient current as a function of time at different temperatures, with $k_BT=0,\Gamma,10\Gamma$, respectively. (c). Contribution to the transient current at zero temperature from m-transferred electrons. (d). Construction of WTD from cumulant expansion.}
 \label{fig1}
\end{figure}

In Fig.~\ref{fig1}b, we plot the transient currents at different temperatures ($k_B T=0$, $\Gamma$, and $10\Gamma$ respectively). At $T=0$ the current rises quickly to the maximum transient current and then shows damped oscillatory behavior in reaching the steady state limit. The steady state dc current can be checked by a separate calculation from Landauer-Buttiker's formula. This oscillatory behavior resembles the classical charging effect. The frequency of transient current oscillation is given by $\Delta_L/2$ which is equivalent to a period of $T_0=1.26$. The damping rate is dominated by the life time of the resonant state of the quantum dot which is about $1/2$. The relaxation time for transient current to reach the steady state is about 8. As we increase the temperature to $k_B T=\Gamma$, the oscillatory behavior is almost gone and the steady state current is less than that at zero temperature with a much shorter relaxation time. At very high temperature $k_BT=10 \Gamma$, the transient current quickly reaches steady state with no oscillation and the dc current is very small. Similar behaviors have been reported in Ref.\onlinecite{foot2} where the bandwidth was varied instead of temperature. From Eqs.(\ref{eq3}) and (\ref{eq4}), we have $I(t) = e\sum_n n dP(n,t)/dt$. In Fig.~\ref{fig1}c, we plot the contribution of $P(n)$ to the transient current for $n=-1,1,2,3,4$. We see that at short times, $P(\pm 1)$ dominates and the transient dynamics can be well described using a few terms of $P(n)$.

\begin{figure}
  \includegraphics[width=3.4in]{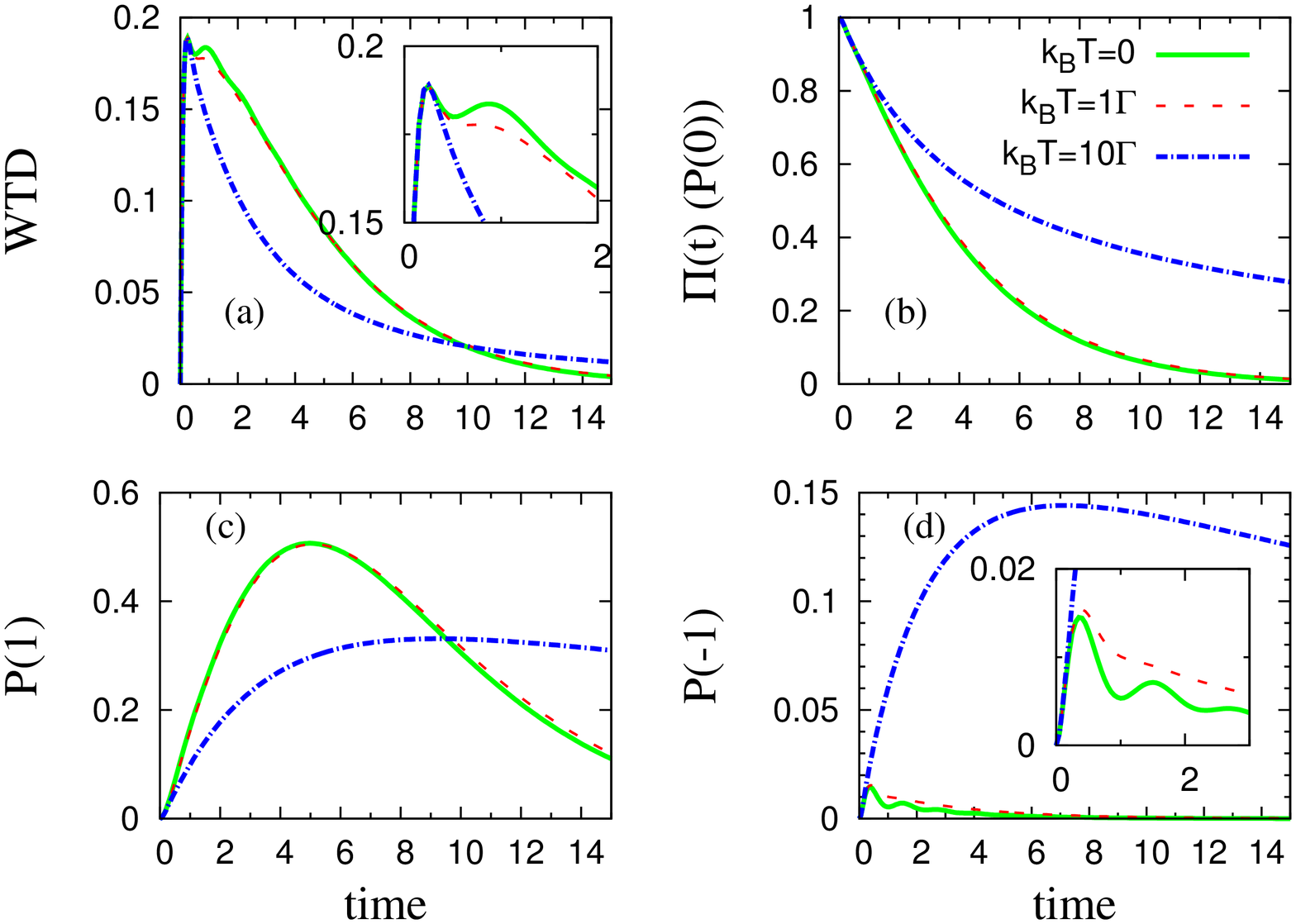}\\
  \caption{waiting time distribution and the probability for zero electron $P(0)$, one electron with positive direction, $P(1)$ and one electron with inverse direction, $P(-1)$ at zero temperature, $k_BT=1\Gamma$ and $k_BT=10\Gamma$ respectively, times are in units of $1/\Gamma$.}
  \label{fig2}
\end{figure}

In Fig.~\ref{fig2}, we present the numerical results of WTD, the probability for detecting zero electron $P(0,t)$ and one electron $P(1,t)$ and $P(-1,t)$ during time interval $t$ at three different temperatures, $k_BT=0,\Gamma,10\Gamma$, respectively. In contrary to the transient current, the WTD and the probability of detecting one electron are not very sensitive to the temperature when temperature is comparable to $\Gamma$, the coupling between leads and quantum dot. For very high temperature $W_1(t)$ decays faster initially and then at a slower rate compared with situations at low temperatures. At long times, the behaviors of $W_1(t)$ at three temperatures follow exponential form $e^{-\kappa t}$ showing Poissonian distribution due to the fact that at long times the scattering events become independent. We notice that WTD at zero temperature has a small oscillation at short times which resembles the charging effect. At short times since the probability for detecting two or more electrons going through the quantum dot is very small, $P(0)$ is approximately equal to $1-P(1)-P(-1)$. We can see from the figure that at short times, $P(-1)$ shows oscillatory behavior that is responsible for the oscillation of $W_1(t)$ as well as transient current at short times. Fig.~\ref{fig2} also shows that at high temperature $k_BT=10\Gamma$, $\Pi(t)$ and $P(-1,t)$ are much larger than that at low temperatures. At $k_BT=10\Gamma$, $P(-1,t)$ does not vanish in the steady state limit and is still very large compared to $P(1,t)$, this explains why the current of $k_BT=10\Gamma$ is much smaller than the low temperature cases.

\begin{figure}
  \includegraphics[width=3.4in]{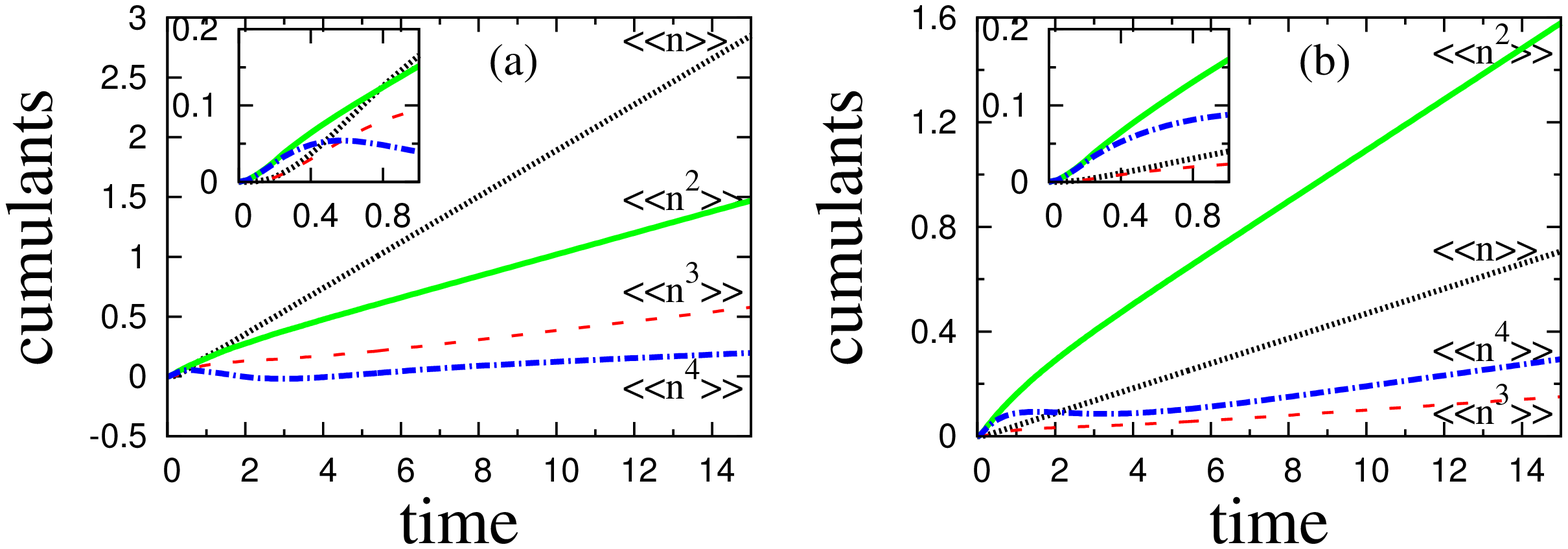}\\
  \caption{Cumulants as a function of time at different temperatures with $T=0$ (panel a) and $k_BT=10\Gamma$ (panel b), respectively.  }
  \label{fig3}
\end{figure}

In Fig.~\ref{fig3}, we present the numerical results for the cumulants as a function of time at zero temperature and $k_BT=10\Gamma$, respectively, from which the linear long time behaviors are clearly seen as a result of Eq.(\ref{eq4}) and (\ref{eq18}). We see from Fig.~\ref{fig3} that at long times $\langle\langle n^{2j} \rangle\rangle$ and $\langle\langle n^{2j+1}\rangle\rangle$ are decreasing functions of $j$. However, this behavior does not hold at short times.

\bigskip

\section{Relation between WTD and cumulants}
Now we discuss how to obtain the WTD from cumulants of transferred charge which can be measured experimentally.\cite{ref1} From Eq.(\ref{eq3}) we can construct a partial sum of CGF $u_m(\lambda,t) \equiv \sum_{j=0}^{m} [(i\lambda)^j/j!] \langle\langle (\Delta n)^j \rangle\rangle$ where only a finite number of experimental measured cumulants are included since the series converges from the observation of Fig.~\ref{fig3}. The approximated WTD can be obtained numerically
\begin{equation} \label{eq21}
W_1^m(t) = - \int_0^{2\pi} \frac{d\lambda}{2\pi} \exp[u_m(\lambda,t)] \partial_t u_m(\lambda,t).
\end{equation}
In Fig.1d, we calculate WTD using Eq.(\ref{eq21}) by including first mth cumulants where $m=8,10,12,14$ in the short times regime where the convergence is the worst. We see that by including more cumulants the approximated $W_1^m(t)$ converges to the exact result. Beyond $t=6$, the $W_1^m(t)$ agrees with $W_1(t)$. Since cumulants of transferred charge have been measured experimentally, the WTD can be obtained using the information of cumulants of transferred charge.

\bigskip

\section{Conclusion}
We have presented a theoretical formalism to investigate FCS and WTD in the transient regime. In this formalism, the GF has been expressed in terms of non-equilibrium Green's function in Keldysh space and can in principle be implemented in the first principles calculation by combining non-equilibrium Green's function with density functional theory. We have applied this theory to a quantum dot coupled with two leads with finite bandwidth and solved Green's functions exactly in the transient regime. This enables us to calculate cumulants of transferred charges, its probability distribution function, and WTD in the transient regime. We analyze short and long time behaviors of WTD as well as the thermal noise contribution to the cumulants and WTD. We have also discussed how to obtain WTD using quantities that can be measured experimentally.

\begin{acknowledgements}
We thank J.S. Wang for useful discussions. This work was financially supported by the Research Grant Council (Grant No. HKU 705212P) and the University Grant Council (Contract No. AoE/P-04/08) of the Government of HKSAR.
\end{acknowledgements}


\begin{thebibliography}{22}
\bibitem{foot1}
According to Ref.1, WTD can be defined in two ways: two-measurement WTD $W_2$ and one measurement WTD $W_1$. Here $W_2$ is used by most people while $W_1$ is the special case suitable for the transient regime. Most of the discussion in the paper is related to $W_1$.
\bibitem{foot4}
The same expression has been obtained in Ref.\onlinecite{21} using superoperators in Liouville space.
\end{thebibliography}
\end{document}


\title{Supplemental material: Waiting Time Distribution of Quantum Electronic Transport in Transient Regime}

\author{Gao-Min Tang}
\author{Fuming Xu}
\author{Jian Wang}
 \email{jianwang@hku.hk}
\affiliation{Department of Physics and the Center of Theoretical and Computational Physics, The University of Hong Kong, Pokfulam Road, Hong Kong, China}

\maketitle
In this supplemental material we present details on how to calculate Green's function and self-energy in the time domain in the presence of an upward pulse.\cite{35} The equilibrium self-energies are chosen to be energy dependent with a finite band width $W_0$
\begin{equation}  \label{eq1}
\tilde{\Sigma}_{\alpha}^r(\omega)=\frac{\Gamma_{\alpha} W_0}{2(\omega+iW_0)} \tag{1}
\end{equation}
so that the linewidth function is the following Lorentzian form
\begin{equation}  \label{eq2}
{\bf \Gamma}_{\alpha}(\epsilon)=\frac{\Gamma_{\alpha} W_0^2}{\epsilon^2+W_0^2}  \tag{2}
\end{equation}
where $\Gamma_{\alpha}$ is the linewidth amplitude.
The self-energy in the time domain is defined as
\begin{equation}   \label{eq3}
\Sigma_{\beta}^{r,<}(\tau_1,\tau_2)=\int \frac{d\omega}{2\pi} e^{-i\omega(\tau_1-\tau_2)}\tilde{\Sigma}_{\beta}^{r,<}(\omega) e^{ -i\int_{\tau_2}^{\tau_1}\Delta_L(t) dt }   \tag{3}
\end{equation}
where $\tilde{\Sigma}_{\beta}^{r,<}$ is the equilibrium self-energy in the energy domain. Using Eq.(\ref{eq3}), we find the retarded self-energy of the left lead:
\begin{equation}   \label{eq4}
\Sigma_L^r(\tau_1,\tau_2)=-\frac{i}{4}
\theta(\tau_1-\tau_2)\Gamma W_0 e^{-(i\Delta_L+W_0)(\tau_1-\tau_2)}   \tag{4}
\end{equation}
where we have assumed $\Gamma_L=\Gamma_R=\Gamma/2$.
For the lesser self-energy
\begin{equation}    \label{eq5}
\Sigma_L^<(\tau_1,\tau_2)=i\int\frac{d\omega}{2\pi} e^{-i\omega(\tau_1-\tau_2)}
e^{-i\Delta_L(\tau_1 -\tau_2)}f(\omega){\bf \Gamma}_L(\omega)   \tag{5}
\end{equation}
with $f(\omega)=1/\left[e^{\beta(\omega-E_F)}+1\right]$.

At zero temperature, we have
\begin{align}    \label{eq6}
\Sigma_L^<(\tau_1,\tau_2)
&=i e^{-i\Delta_L(\tau_1 -\tau_2)} \int_{-\infty}^{0}\frac{d\omega}{2\pi} e^{-i\omega(\tau_1-\tau_2)}\frac{\Gamma_L W_0^2}{\omega^2+W_0^2}   \tag{6}
\end{align}
where we have set $E_F=0$.  \\
1. If $\tau_1=\tau_2$
\begin{equation}    \label{eq7}
\Sigma_L^<(\tau_1,\tau_2)=\frac{i}{8}\Gamma W_0    \tag{7}
\end{equation}
2. If $\tau_1>\tau_2$, let $\tau=\tau_1-\tau_2$
\begin{align}    \label{eq8}
\Sigma_L^<(\tau_1,\tau_2) &=\frac{i}{8}\Gamma W_0
\left\{\frac{i}{\pi} e^{(W_0-i\Delta_L)\tau} E1(W_0\tau)  \right. \notag \\
&\left.  +e^{-(W_0+i\Delta_L)\tau} \left[2-\frac{i}{\pi}E1(-W_0\tau) \right]\right\}
 \tag{8}
\end{align}
where $E1(x)=\int_x^{\infty}\frac{e^{-t}}{t}dt$.

At non-zero temperature, we have

1. if $\tau_1=\tau_2$, the integral is actually Hilbert transformation of the Fermi distribution function.\cite{36}
\begin{equation}   \label{eq9}
\Sigma_L^<(\tau_1,\tau_2)=\frac{i\Gamma W_0}{8}   \tag{9}
\end{equation}
2. if $\tau_1>\tau_2$, it has poles $\frac{-i(2n+1)\pi}{\beta}$ and $-iW_0$, where $n=0,1,2,3...$
\begin{align}  \label{eq10}
&\Sigma_L^<(\tau_1,\tau_2)=
\frac{i\Gamma_L W_0\exp[-(W_0+i\Delta_L)(\tau_1-\tau_2)]}{2\exp(-i\beta W_0)+2}
-\frac{1}{\beta} \times    \notag \\
&\sum_{n=0}^{+\infty}\exp\left\{-[\frac{(2n+1)\pi}{\beta}+i\Delta_L](\tau_1-\tau_2) \right\}\frac{\Gamma_L W_0^2}{W_0^2-[\frac{(2n+1)\pi}{\beta}]^2}    \tag{10}
\end{align}
Using the relation $\Sigma_L^<(\tau_1,\tau_2)\big|_{\tau_1<\tau_2}=-[\Sigma_L^<(\tau_1,\tau_2)\big|_{\tau_1>\tau_2}]^*$, we obtain the full expression of $\Sigma_L^<(\tau_1,\tau_2)$. The expression of $\Sigma_R^<(\tau_1,\tau_2)$ can be obtained similarly.

Taking Fourier transform of the retarded self-energy $\Sigma^r(\tau_1,\tau_2)=\Sigma_L^r(\tau_1,\tau_2)+\Sigma_R^r(\tau_1,\tau_2)$ to energy domain, we find
\begin{equation}  \label{eq11}
\Sigma^r(\omega)=\frac{\Gamma W_0}{4}[\frac{1}{\omega+iW_0}+\frac{1}{\omega-\Delta_L+iW_0}]    \tag{11}
\end{equation}
The retarded Green's function in time domain is given by
\begin{align}  \label{eq12}
G^r(\tau_1,\tau_2)&=\int\frac{d\omega}{2\pi}e^{-i\omega(\tau_1-\tau_2)}\frac{1}{\omega-\epsilon_0-\Sigma^r(\omega)}   \notag  \\
&=\int\frac{d\omega}{2\pi}e^{-i\omega(\tau_1-\tau_2)}\frac{(\omega+iW_0)(\omega-\Delta_L+iW_0)}{(\omega-\omega_1)(\omega-\omega_2)(\omega-\omega_3)}
\tag{12}
\end{align}
 where $\omega_1, \ \omega_2, \ \omega_3$ are poles of equilibrium retarded Green's function satisfying the following equation
\begin{equation}  \label{eq13}
\omega^3 + a \omega^2 +b\omega +c=0  \tag{13}
\end{equation}
with
\begin{align} \label{eq14}
a &=(2iW_0-\Delta_L-\epsilon_0)  \notag \\
b &=iW_0(iW_0+\frac{i\Gamma}{2}-2\epsilon_0-\Delta_L)+\epsilon_0\Delta_L  \notag \\
c &=\frac{\Gamma W_0}{4} (\Delta_L-2iW_0)+iW_0\epsilon_0(\Delta_L-iW_0)   \tag{14}
\end{align}
This integral can be calculated using the theorem of residue.
\begin{equation}  \label{eq15}
G^r(\tau_1,\tau_2)=-i\sum_i^3\frac{(\omega_i+iW_0)(\omega_i-\Delta_L+iW_0)e^{-i\omega_i(\tau_1-\tau_2)}}{\sum_{j,k=1}^3|\epsilon_{ijk}|(\omega_i-\omega_j)(\omega_i-\omega_k)}   \tag{15}
\end{equation}
where $\epsilon_{ijk}$ is the usual Levi-Civita symbol. Here $G^<$ can be calculated through the Keldysh equation:
\begin{equation}   \label{eq16}
G^<=(1+G^r\Sigma^r)g^<(1+\Sigma^aG^a)+G^r\Sigma^<G^a   \tag{16}
\end{equation}
The first term is zero if $(g^r)^{-1}g^< (g^a)^{-1}=0$ which is the case for steady states. In our case we should keep it in the transient regime calculation. In the calculation, we have chosen $g^<=0.5i$ at $t=0$ which makes sure that $I_L+I_R=0$ and $<(\Delta I_L)^2>=<(\Delta I_R)^2>$.